# TMOKE as efficient tool for the magneto-optic analysis of ultra-thin magnetic films

O.V. Borovkova[1], H. Hashim[2,3], M.A. Kozhaev[1,4], S.A. Dagesyan[7], A. Chakravarty[5,6], M. Levy[5], and V.I. Belotelov[1,7]

[1]*Russian Quantum Center, Skolkovo, Moscow Region 143025, Russia*
[2]*National University of Science and Technology (MISIS), Moscow 119049, Russia*
[3]*Physics Department, Faculty of Science, Tanta University, Tanta 31527, Egypt*
[4]*Prokhorov General Physics Institute RAS, Moscow 119991, Russia*
[5]*Physics Department, Michigan Technological University, Houghton, Michigan 49931, USA*
[6]*Radboud University, Nijmegen 6525 HP, Netherlands*
[7]*Lomonosov Moscow State University, Moscow 119991, Russia*

Ultra-thin magnetic dielectric films are of prime importance due to their applications for nanophotonics and spintronics. Here we propose an efficient method for the magneto-optical investigation of ultra-thin magnetic films that allows one to access their state of magnetization and magneto-optical properties. It is based on the surface-plasmon-polariton-assisted transverse magneto-optical Kerr effect (TMOKE). In our experiments sub-100nm-thick bismuth-substituted lutetium iron-garnet films covered with a plasmonic gold grating have been analyzed. The excitation of surface plasmon-polaritons provides resonance enhancement of TMOKE up to 0.04 and makes it easily detectable in experiment. For films thicker than 40 nm the TMOKE marginally depends on the film thickness. Further decrease of the film thickness diminishes TMOKE since for such thicknesses the surface plasmon-polariton field partly penetrates inside the nonmagnetic substrate. Nevertheless, the TMOKE remains measurable even for few-nm-thick films, which makes this technique unique for the magneto-optical study of ultra-thin films. Particularly, the proposed method reveals that the off-diagonal components of the magnetic film permittivity tensor grow slightly with the reduction of the film thickness.

Currently, ultra-thin ferrimagnetic dielectric films are of significant interest due to their applications in nanophotonics, magnonics and spintronics [1-6]. Magnetic dielectrics like bismuth-substituted iron-garnets have outstanding optical properties in the near infrared where they evince low optical absorption and a relatively large magneto-optical response [7-10]. Practical use of spintronic devices for magnetic information recording requires magnetic field confinement in the nanoscale regime [1-4]. Hybrid structures like iron-garnet/platinum and iron-garnet/topological insulators which are considered as very promising for spin current generation and control demand nm-thick magnetic dielectric films in order to enhance the impact from surface related effects like the Rashba-Edelstein and Spin-Hall effects [11-12]. Magnonic devices currently under intense research also require ultra-thin insulating films to achieve desired levels of miniaturization and efficiency [13]. In addition, such ultra-thin films with a thickness of tens of nanometers are advantageous for sensing applications as an extremely large proportion of surface atoms that serve as adsorption centers and the high field confinement in such films enable higher sensitivity [1,5].

The magneto-optical response of the ultra-thin films evinces some peculiarities. In particular, the specific Faraday rotation in iron-garnet films has been found to grow as the thickness decreases down to tens of nanometers [14]. Such increase might be due to a larger magneto-optical gyrotropy parameter near the film interface than in the bulk. It should be noted that the Faraday effect grows with the optical path-length in the sample. For that reason, a Faraday rotation acquired from the whole magnetic film drops for thinner films and becomes hardly detectable when the film thickness is just a few nanometers. In this respect the transverse magneto-optical Kerr effect (TMOKE) is promising as it is highly sensitive to the magnetization near the sample surface and therefore can sustain decent values even for ultra-thin films [15]. Moreover, its inverse counterpart is of prime importance for the ultrafast magnetism [16]. However, TMOKE appears for absorbing media and is very small for transparent dielectrics.

If a magnetic film is covered with metal film, enabling the excitation of surface plasmon-polaritons (SPP), the magneto-optical effects can be significantly enhanced near the SPP resonance and the TMOKE could be as high as tens of a percent [17-22]. This makes it amenable for the investigation of the transparent magneto-optical films.

So far, TMOKE has been observed only for metal covered thick magnetic films of thicknesses greater than 100 nm. However, TMOKE is marginally sensitive to the bulk properties and is mainly governed by the optical properties near the magnetized interface. Therefore, it is advantageous to investigate TMOKE for magnetoplasmonic structures based on much thinner films.

In this Letter we investigate TMOKE in bismuth-substituted iron-garnet films of thickness less than 60 nm with a one-dimensional gold grating cover. We analyze the plasmonically enhanced TMOKE peculiarities appearing when the film thickness is decreased. It is found that they can be explained by a variation of the non-diagonal element of the dielectric tensor and by localization of the SPPs at the metal-dielectric interface.

For the experimental studies we use $Bi_{0.8}Gd_{0.2}Lu_2Fe_5O_{12}$ films of thicknesses 19 nm, 46 nm, and 60 nm. These samples were grown by liquid-phase epitaxy (LPE) on (100) gadolinium gallium garnet ($Gd_3Ga_5O_{12}$) substrates (GGG). The films exhibit planar magnetic anisotropy and saturation magnetization of $4\pi M_s$=1800 G. All the samples were obtained from the same original magnetic film by a sequential etching in an ortho-phosphoric acid bath with a slow-rotation rate to ensure uniform thickness [14]. Thickness uniformity was verified by probing different points of the sample surface via transmission electron microscopy.

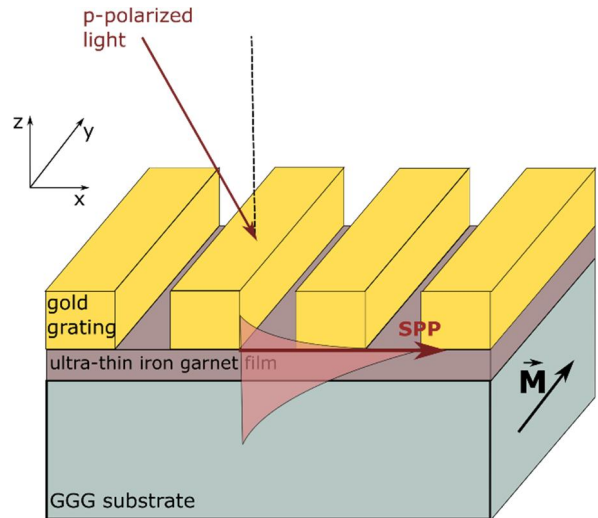

Figure 1. The scheme of the plasmonic crystal. Ultra-thin LPE-grown bismuth-substituted iron garnet film on GGG substrate and covered by sub-wavelength gold grating is illuminated by obliquely incident p-polarized light.

To fabricate the magnetoplasmonic crystal (Fig. 1) an 80nm-thick gold layer was deposited on the magnetic films by magnetron sputtering. A resistive mask was then formed on top via electron-beam lithography, and gold-layer patterning was performed by ion etching in an argon-ion plasma single-frequency discharge. The pattern is a periodic array of slits and its period was determined by preliminary numerical modelling to provide the extrema of TMOKE resonances at the same frequencies for the different samples. This allows one to avoid the dielectric dispersion impact and to isolate the influence of the magnetic garnet film thickness on the magneto-optical response.

The parameters of the gratings were determined from the electromagnetic modelling based on a rigorous coupled-waves analysis (RCWA) [23, 24]. Varying both the grating period and the slits width, we selected the optimal air slit width of $r$ = 75nm to maximize the TMOKE peak for all samples. In its turn, the grating period determines the resonance position. Emergence of TMOKE resonances at the same frequency allows one to eliminate the impact of the material-parameter dispersion and to isolate the influence of the magnetic garnet film thickness on the magneto-optical response. The grating period values were found to be $d$ = 318nm for the 60nm-thick film, $d$ = 327nm for the 46nm-thick film, and $d$ = 350nm for the 19nm-thick film.

Naturally, the parameters of the fabricated gratings evince some deviations from the optimal parameters, as determined by electron microscopy. In particular, the actual plasmonic grating periods and widths of the slits were found to be $d$ = 324nm, $r$ = 85nm for the 60nm-thick film, $d$ = 322nm, $r$ = 75nm for the 46nm-thick film, and $d$ = 347nm, $r$ = 60nm for the 19nm-thick film. Nevertheless, as required, all three samples demonstrate pronounced TMOKE resonances in the same wavelength range.

The experimental setup we used allows us to measure magneto-optical effects over a wide range of wavelengths and incidence angles. A tungsten halogen lamp is used as light source in the visible and near-IR ranges. The light is collimated with an achromatic lens (with a focal distance 75 mm) and focused onto the sample with another achromatic lens (with a focal distance 35 mm) into a spot of ~200 μm. The sample is placed in a uniform external magnetic field of 2000 Oe along *y*-axis in Fig. 1 generated by the electromagnet. The applied magnetic field exceeds the magnetic field required to saturate the magnetization of the iron garnet films under consideration and guarantees the reproducibility of the results. The light is collimated after it exits the sample with a 20x microscope objective and detected with the spectrometer. The spectrometer slit is oriented perpendicularly to the sample gratings slits, so only the light with incidence in the *xy* plane in Fig. 1 is detected. A 2D CCD camera in the spectrometer is used to observe the spectral decomposition along one axis and the incidence angle decomposition along the perpendicular axis. Therefore, we measure the angular and wavelength resolved transmission spectra of all samples for two opposite directions in the magnetic field. Each measurement with alternating opposite directions of the magnetic field is repeated 200 times and then these results are averaged. This regime provides reproducibility of the measurements, with a signal to noise ratio exceeding three orders of magnitude in the spectral range of our interest. Based on these spectra we can find δ, the value of the TMOKE, as a relative change of the transmitted light intensity *T(M)* when the structure is re-magnetized:

$$\delta = 2\frac{T(\boldsymbol{M})-T(-\boldsymbol{M})}{T(\boldsymbol{M})+T(-\boldsymbol{M})}. \tag{1}$$

The measured wavelength-, and angular-resolved transmission (left column), and TMOKE (right column) spectra of the three samples are given in Fig. 2. Excitation of SPPs at the [gold]/[ferromagnetic dielectric] interface leads to the dips in the transmission spectra as follows from the calculation of the SPP dispersion based on the phase synchronism condition (dashed lines in Fig. 2) [25]. The corresponding calculated transmission and TMOKE spectra are given in the Supplementary Material. A second-, and third-band SPPs are observed. The transverse in-plane magnetic field spectrally shifts the transmission dips either towards lower or higher frequencies depending on the direction of the magnetic field with respect to the normal to the sample surface and SPP wavevector. As a result, near the frequencies of SPP, excitation resonances in the TMOKE spectra appear. They have an S-shape with positive and negative maxima where the TMOKE reaches 0.04. The TMOKE spectra is antisymmetric with respect to the normal incidence when TMOKE vanishes due to symmetry reasons. Although the thicknesses of the magnetic garnet films in the plasmonic crystals under consideration are different from each other, the resonance positions in all three cases are almost the same due to the proper choice of the grating periods.

If the magnetic garnet films are thick enough then the TMOKE hardly depends on the film thickness (compare the TMOKE spectra for the 60-nm and 46-nm thick films in Fig. 1b,d). However, for thinner films the TMOKE tends to decrease. Nevertheless, the decrease is not very pronounced so that the TMOKE maximum drops to 0.03 for the 19-nm thick film (Fig. 1f). Such behavior in the TMOKE can be due to the changes in the magnetic garnet film permittivity-tensor and modifications of the SPP modes caused by the decrease in film thickness.

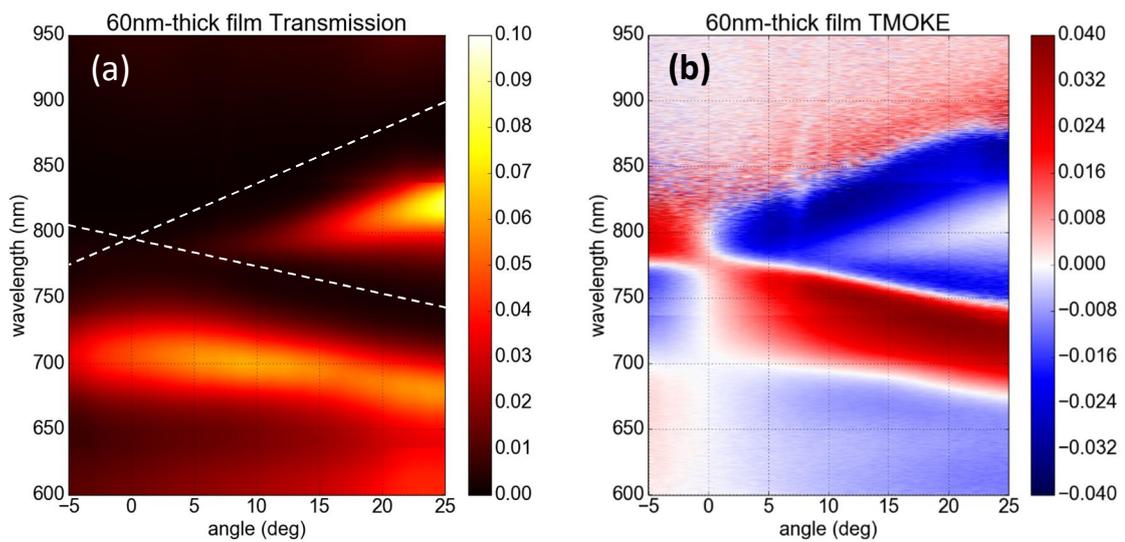

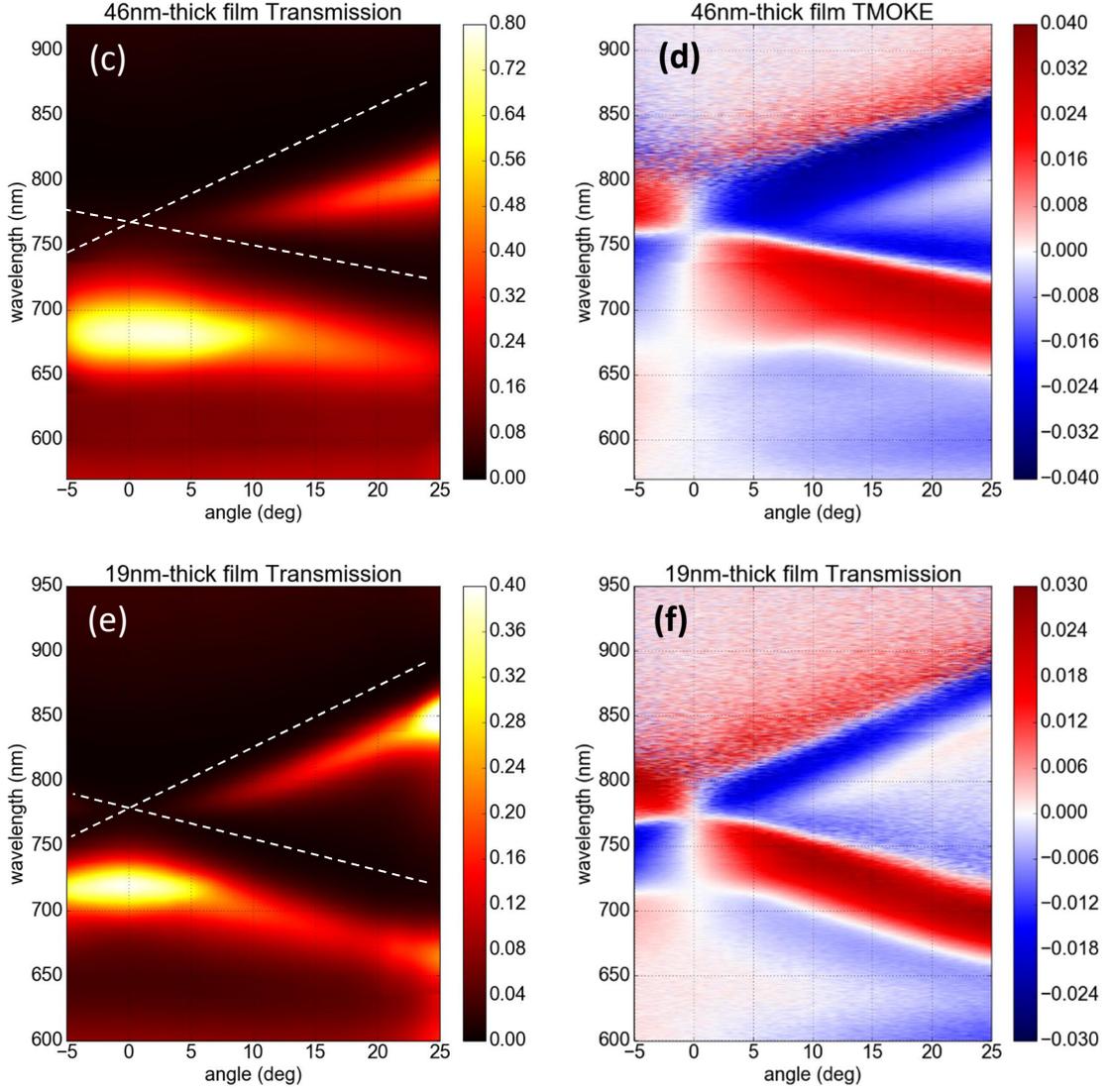

Figure 2. Wavelength and angular resolved transmission (a, c, and e) and TMOKE (b, d, and f) spectra of the 60nm-thick- (a-b), 46nm-thick, (c-d), and 19nm-thick-bismuth-substituted iron garnet films (e-f). White dashed lines show the SPP dispersion calculated in the simplified model of smooth interfaces. Gold grating periods are 324nm (60nm-thick film), 322nm (46nm-thick film), and 347nm (19nm-thick film).

To investigate the physical origin of the observed TMOKE behavior we perform an electromagnetic modelling based on rigorous coupled-waves analysis (RCWA) [23, 24]. The refractive indices of gold, GGG, and bismuth-substituted iron garnet are taken from [26, 27]. Geometrical parameters of the gold grating are measured by TEM imaging. There remains only one unknown parameter: the off-diagonal components of the magnetic garnet film permittivity tensor, $\varepsilon_1$. Therefore, $\varepsilon_1$ can be found by matching the calculated and experimentally measured transmission and TMOKE spectra.

Let us consider in detail the spectral range between 0.7 and 0.9 μm in wavelength where two TMOKE S-shape resonances are present (dots, Fig. 3). These are caused by SPPs propagating in opposite directions which makes their signs opposite to each other. At 10° incidence the two resonances are relatively close so that a slight overlap between them takes place. For a 20° incidence the resonances get separated enough and do not interference with one another.

For the calculations of TMOKE in the spectral range under consideration we must take into account that the dispersion in $\varepsilon_1$ is determined by the optical transitions in the $Fe^{3+}$ ions in the octahedral- and tetrahedral-coordinated sub-lattices of the iron garnet and can be described by the expression from [28, 29] (Supplementary). The calculated TMOKE spectral curves show good agreement with the measured ones: they are identical in shape but there is a small discrepancy in values that might be due to departures in the gold grating cross-section profile from the assumed rectangular one (solid curves, Fig. 3 and calculated 2D transmission and TMOKE spectra are given in Fig. S1 in the Supplementary Material).

It should be noted that $\varepsilon_1$ found for all three samples have similar dispersion but are different in values. Particularly, at $\lambda = 0.754$ μm, close to one of the TMOKE resonances, $\varepsilon_1 = 0.0084$ for the 60-nm-thick film, $\varepsilon_1 = 0.0088$ for the 45-nm-thick film, and $\varepsilon_1 = 0.0102$ for the 19-nm-thick film. Therefore, TMOKE measurements indicate a slight growth in the magneto-optical gyrotropy parameter of the film as the thickness decreases. This agrees with the previous results obtained from Faraday effect measurements [14].

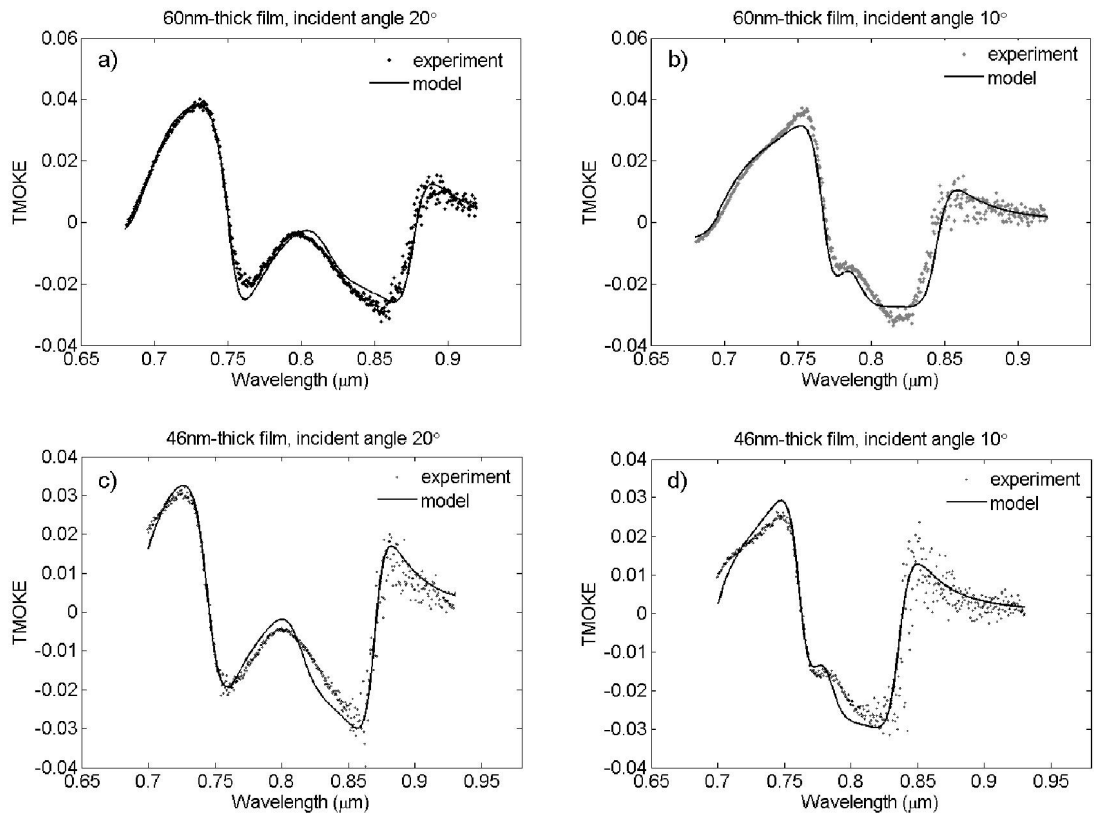

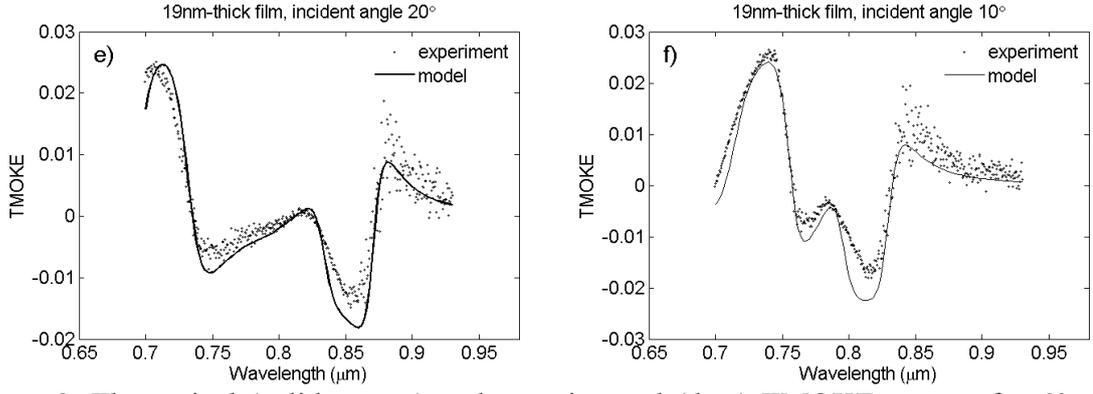

Figure 3. Theoretical (solid curves) and experimental (dots) TMOKE spectra for 60nm-thick films (a, b), 46nm-thick films (c, d), and 19nm-thick films (e, f) for 20º and 10 º incident angles. Gold grating periods are 324nm (60nm-thick film), 322nm (46nm-thick film), and 347nm (19nm-thick film).

Interestingly, in spite of the growth of $\varepsilon_1$ the TMOKE decreases for thinner films. As TMOKE is related to the plasmonic resonances, this phenomenon can be understood by analyzing the SPP modes in these structures. The spatial distribution of the electromagnetic field at the SPP resonance (Fig. 4) indicates that the penetration depth of the SPP wave is comparable to the thickness of the magnetic garnet films in the samples. The SPP field amplitude decreases by a factor of *e* at the depth of about 40 nm. In plasmonic crystals with 60nm-thick and 46nm-thick films SPP field is localized mostly inside the ultra-thin magnetic garnet film. However, in the 19nm-thick film a significant part of the SPP field penetrates into the non-magnetic GGG substrate. This diminishes the influence of the magnetic field on the SPP in the 19nm-thick film sample, the SPP frequency is shifted by a smaller value and, consequently, the TMOKE decreases even though $\varepsilon_1$ is a bit larger.

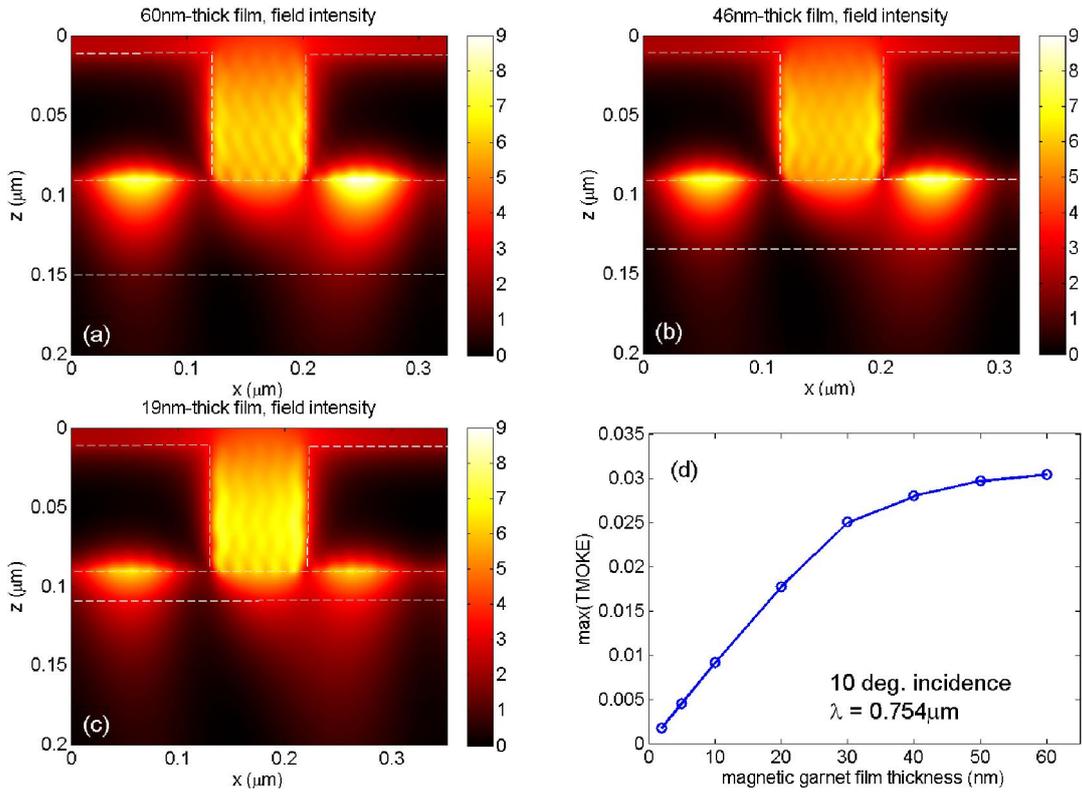

Figure 4. The SPP wave field distribution in the plasmonic crystal with (a) 60nm-thick-, (b) 46nm-thick-, and (c) 19nm-thick-bismuth-susbituted iron garnet film. White dashed lines mark the contours of the gold grating and the ferromagnetic film. Gold grating periods are 324nm (60nm-thick film), 322nm (46nm-thick film), and 347nm (19nm-thick film). Incidence angle is 20º. (d) Calculated dependence of the TMOKE maximum value at $\lambda = 0.754$ μm on the magnetic film thickness.

If $\varepsilon_1$ is assumed to be independent of film thickness, then the TMOKE decrease in the thinner films is more pronounced as follows from the calculated curve in Fig. 4d. Further decrease in film thickness makes the TMOKE smaller but it has relatively large values even for the nm-thick films. For example, for a 2-nm-thick film sample the TMOKE is $1.8 \cdot 10^{-3}$ which can be easily detectable. It should be noted that the Faraday effect for such films is very small ($2.7 \cdot 10^{-6}$ deg. for the 2-nm-thick film) and hardly measurable.

As it has been shown previously, a proper plasmonic cover also enhances the Faraday effect [30, 31], but this approach does not work for ultra-thin films. The reason is that the Faraday effect is enhanced for waveguide modes but because of the cut-off frequency these modes appear in such thin films only in the short-wavelength visible range regime where iron-garnets are highly absorptive. Consequently, plasmonic TMOKE is a unique technique sensitive to the magnetization of nm-thick magnetic dielectric films.

Although, the proposed method of the magneto-optic analysis of ultra-thin magnetic films by the SPP-enhanced TMOKE employs a resonant effect, it still admits the spectral detuning simply by a change of the incidence angle. As one can see in Figs. 2-3 SPP-enhanced TMOKE resonance has an S-shape with two extrema of opposite signs. The spectral width of one extremum is about 50nm. Varying the incidence angle from 0 to 25 degrees one can observe SPP-enhanced TMOKE resonances (corresponding to second-, and third-band SPPs) in the spectral range from 700nm to 900nm. Thus, the angularly resolved measurements in the proposed method provide the spectroscopic information of the magneto-optical properties of ultra-thin magnetic films even though the method employs a resonant effect.

Moreover, to secure a wider spectral range in the measurements for an unknown sample, a set of gold gratings with the various values for the period can be used. For instance, if the grating period is 20nm greater with all other parameters taken the same, one gets the spectral range of SPP resonances shifted approximately by 40nm with respect to the initial spectral range.

To conclude, in this paper we address the transverse magneto-optical Kerr effect in ultra-thin bismuth-substituted iron garnets covered with plasmonic gratings. The plasmonic structure provides resonance in the TMOKE spectrum when it reaches 0.04 for the samples with 60nm-thick and 46nm-thick films. However, the sample with a thinner magnetic garnet film, a 19nm-thick one, gives smaller TMOKE signal of 0.03 at the most. Theoretical analysis of the TMOKE spectra for all samples based on the magneto-optical model taking into account a sum of the $Fe^{3+}$ sub-lattice transitions shows that the observed TMOKE spectra are governed by two factors. On the one hand, the off-diagonal component of the permittivity tensor of the magnetic film grows slightly with reduced film thickness. On the other hand, the electromagnetic field distribution in the SPP wave also affects the signal. In particular, for thinner films the field of the SPP wave partially penetrates the non-magnetic substrate which weakens the influence of the magnetic field on it and lowers the TMOKE. Nevertheless, calculations demonstrate that TMOKE remains relatively large even for few-nm-thick films in contrast to the Faraday effect, which becomes negligibly small and hardly detectable at these scales. Therefore, TMOKE in ultra-thin films with plasmonic cover is a unique technique and very advantageous for the investigation of ultra-thin magnetic films, their characterization and their applications in nanophotonic and spintronic devices.

**Supplementary material**

See supplementary material for the iron garnet gyrotropy dispersion and calculated transmission and TMOKE spectra.

**Acknowledgements**

This study was supported by Russian Foundation for Basic Research (project no. 16-32-60135 mol_a_dk), and by Foundation for the advancement of theoretical physics BASIS. The work was carried out also with financial support from the Ministry of Education and Science of the Russian Federation in the framework of increase Competitiveness Program of NUST 'MISIS', implemented by a governmental decree dated 16th of March 2013, No. 211.